\documentclass[aps,nature,twocolumn,longbibliography,superscriptaddress,floatfix,nofootinbib]{revtex4-2}
\usepackage{epsfig,amsmath,amssymb,color,comment,physics}
\usepackage[makeroom]{cancel}
\usepackage[caption=false]{subfig}
\usepackage{mathrsfs}
\usepackage[countmax]{subfloat}
\usepackage[normalem]{ulem}
\usepackage[english]{babel}
\usepackage{dsfont}
\usepackage{float}
\usepackage[bookmarks=true,colorlinks,linkcolor=black,urlcolor=NavyBlue,citecolor=RoyalBlue]{hyperref}
\usepackage[dvipsnames]{xcolor}
\usepackage{mathtools}
\usepackage{siunitx}
\usepackage{upgreek}
\usepackage{comment}
\usepackage{mathtools}

\usepackage[normalem]{ulem}

\begin{document}

\newcommand{\titleinfo}{Observing dissipationless flow of an impurity in a strongly repulsive quantum fluid}
\title{\titleinfo}

\author{ Milena Horvath}
\thanks{These authors contributed equally to this work.}
\affiliation{Institut f{\"u}r Experimentalphysik und Zentrum f{\"u}r Quantenphysik, Universit{\"a}t Innsbruck, Technikerstra{\ss}e 25, Innsbruck, 6020, Austria}

\author{ Sudipta Dhar}
\thanks{These authors contributed equally to this work.}
\affiliation{Institut f{\"u}r Experimentalphysik und Zentrum f{\"u}r Quantenphysik, Universit{\"a}t Innsbruck, Technikerstra{\ss}e 25, Innsbruck, 6020, Austria}

\author{ Elisabeth Wybo}
\affiliation{Technical University of Munich, TUM School of Natural Sciences, Physics Department, 85748 Garching, Germany}
\affiliation{Munich Center for Quantum Science and Technology (MCQST), Schellingstr. 4, 80799 M{\"u}nchen, Germany}

\author{Dimitrios Trypogeorgos}
\affiliation{Institute of Nanotechnology, Consiglio Nazionale delle Ricerche (CNR-Nanotec), via Monteroni 165, 73100, Lecce, Italy} 

\author{ Yanliang  Guo }
\affiliation{Institut f{\"u}r Experimentalphysik und Zentrum f{\"u}r Quantenphysik, Universit{\"a}t Innsbruck, Technikerstra{\ss}e 25, Innsbruck, 6020, Austria}
\affiliation{Key Laboratory of Quantum State Construction and Manipulation (Ministry of Education), School of Physics, Renmin University of China, Beijing 100872, China}

\author{ Mikhail Zvonarev}
\affiliation{Université Paris-Saclay, CNRS, LPTMS, 91405, Orsay, France}

\author{ Michael Knap}
\affiliation{Technical University of Munich, TUM School of Natural Sciences, Physics Department, 85748 Garching, Germany}
\affiliation{Munich Center for Quantum Science and Technology (MCQST), Schellingstr. 4, 80799 M{\"u}nchen, Germany}

\author{ Manuele  Landini}
\affiliation{Institut f{\"u}r Experimentalphysik und Zentrum f{\"u}r Quantenphysik, Universit{\"a}t Innsbruck, Technikerstra{\ss}e 25, Innsbruck, 6020, Austria}

\author{ Hanns-Christoph  N{\"a}gerl}\email{christoph.naegerl@uibk.ac.at}
\affiliation{Institut f{\"u}r Experimentalphysik und Zentrum f{\"u}r Quantenphysik, Universit{\"a}t Innsbruck, Technikerstra{\ss}e 25, Innsbruck, 6020, Austria}

\begin{abstract}
The frictionless motion of an object through a fluid medium is commonly viewed as a hallmark of superfluidity. According to Landau, kinematic constraints prohibit superfluid behavior in one-dimensional (1D) bosonic systems. Here, using ultracold atoms, we show how a microscopic impurity can propagate through a strongly interacting 1D Bose gas without any friction, at odds with conventional expectations. We inject the impurity with initial velocities ranging from the subsonic to supersonic regime, and subsequently track its dynamics. For  supersonic initial velocities, we observe the formation of a shock wave and a remarkably fast relaxation to a stationary regime, on a time scale that increases with decreasing impurity velocity. After reaching the stationary state, the impurity continues its motion through the system with a finite velocity. Our findings demonstrate how quantum effects can conspire to eliminate dissipation of a microscopic object immersed in a quantum fluid, thereby bringing novel insights into the propagation of matter and information in the quantum realm.

\end{abstract}

\maketitle

\begin{figure}[t!]
\renewcommand{\figurename}{FIG.}
\centering
\includegraphics[trim={0cm 0cm 0cm 0cm},clip,width=\columnwidth]{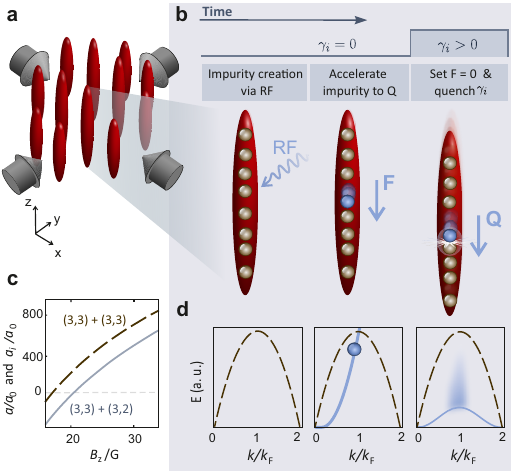}
    	\caption{\textbf{Experimental procedure.} \textbf{a}, Experimental realization of an array of 1D Bose gases in tubes formed by two retro-reflected laser beams. \textbf{b}, Pictorial representation of the experimental sequence. Top panel: interaction strength $\gamma_i$ throughout the experimental sequence.  Middle panel: on average one impurity (blue sphere) per tube is created from a strongly correlated host gas (brown spheres) via a radio-frequency pulse. 
        After its creation the impurity, which is initially not interacting with the host gas, is accelerated by gravity to the desired momentum $Q$. Subsequently the entire system is dropped and interactions between the impurity and the host are switched on. \textbf{c}, Magnetic-field dependence of the scattering length $a$ for collisions between atoms in the host gas (dashed line) and $a_i$ between impurity and host atoms (solid line). \textbf{d}, Left panel: Edge of the excitation spectrum of a 1D Bose gas without impurity -- the plasmon branch (dashed line). Middle panel: The blue parabola corresponds to the dispersion of a single non-interacting impurity. Right panel: The solid line shows the edge of the excitation spectrum with one impurity interacting with the host gas -- the polaron branch. The quench induces excitations (blue shading) in the continuous many-body spectrum above the polaron branch. For both the plasmon and the polaron branch, the excitation spectrum is a $2\,k_\text{F}$-periodic function of the total momentum of the system. } 
	\label{fig:Fig_1}
\end{figure}

Classical mechanics tells us that an impurity moving through a fluid medium inevitably experiences friction as a result of countless collisions with surrounding particles, thus dissipating energy and momentum until it comes to rest~\cite{landau2013fluid}. The laws of quantum mechanics are less restrictive in that regard. A fluid can be in a phase that has no classical analog; this is manifested by the possibility to maintain dissipationless flow when passing a macroscopic obstacle. First observed in liquid helium-4 at extremely low temperatures, this phase is called a superfluid~\cite{Kapitza1938Viscosity,Allen1938Flow}. In 1941, Landau gave a criterion for the existence of such dissipationless flow by considering the collective response to the motion of an external object~\cite{Landau1941Theory}. If the object moves with a velocity $v$ below some critical value $v_c$, elementary excitations cannot be created due to the non-trivial dispersion relation of the superfluid. During the past eight decades, this criterion has underpinned our understanding of superfluidity, guiding both theoretical developments~\cite{Leggett1999,pitaevskii2016bose} and experimental tests ranging from vortex nucleation~\cite{donnelly1991quantized,Matthews1999,madison2000vortex,zwierlein2005vortices} to drag measurements in ultracold quantum gases~\cite{Raman1999,Chikkatur2000,Desbuquois2012}.

Landau's criterion assumes that the impurity is a macroscopic object, sufficiently massive to be treated classically. However, an open question is how this criterion applies in the case of a microscopic object, \textit{i.e.,} a single quantum impurity. Recent experimental advances in local control of ultracold atoms have provided valuable insights into impurity and polaron physics~\cite{Massignan2014,Grusdt2025impurities,Hu2016bosepolaron,Jorgensen2016bosepolaron,Schirotzek2009Fermipolaron,Koschorreck2012,Kohstall2012,Cetina2016,meinert2017bloch,Skou2021noneqPolaron,Baroni2024,Dhar2025}. The quantum nature of the impurity could constrain energy-momentum exchange, allowing frictionless transport beyond the conventional Landau paradigm. A striking example is given for a 1D interacting Bose gas. Since in this system the dispersion dips down to zero energy at finite momentum, the Landau criterion predicts a vanishing critical velocity $v_{c}\!=\!0$~\cite{giamarchi_book_1d}. Therefore, following this reasoning, any mobile macroscopic object should eventually come to rest~\cite{SM,Astrakharchik2004,Cherny2012}. However, recent theoretical work has shown that a quantum impurity can propagate through the gas without dissipation~\cite{mathy2012quantum,Knap2014Flutter,Burovski2014,Lychkovskiy2015,Gamayon2018EndVel}. Addressing this problem in experiments requires the preparation of a finite-velocity initial state of an impurity and probing its relaxation pathway and interaction-driven dressing in real time, which has remained an experimental challenge.

Here, using ultracold Cesium ($^{133}$Cs) atoms, we investigate the motion of an impurity injected with finite velocity into a strongly interacting 1D Bose gas. We initially prepare a non-interacting impurity with a velocity ranging from the subsonic to the supersonic regime. We then quench the system to the strongly repulsive regime and study its relaxation dynamics. We follow the impurity's temporal evolution through the whole relaxation process until it reaches a steady state. In the supersonic regime, we observe the emergence of shock waves and the relaxation occurs within a Fermi time, the fastest possible collective response time of the system. Furthermore, contrary to conventional expectations, we find that in the long-time limit, the impurity never comes to a stop. Instead, it forms a strongly correlated state with the quantum fluid and reaches a steady state that propagates at a reduced velocity.

We consider strongly interacting 1D bosons with mass $m$, prepared in an array of tubes formed by a pair of retro-reflected laser beams, see Fig.~\ref{fig:Fig_1}{\bf a}. Each 1D system is characterized by a dimensionless interaction parameter $\gamma$~\cite{OlshaniiAtomic1998}. Experiments in the strong interaction regime, $\gamma\gg1$, exhibit transmutation of quantum statistics~\cite{Paredes2004, TG_weiss, Haller2009Realization}: the excitation spectrum of the system becomes identical to that of the polarized Fermi gas~\cite{TG}. The time scale governing interparticle collisions is thus determined by the Fermi time, $t_\text{F}\!=\!2m/(\hbar k_\text{F}^2)$, where $\hbar$ is Planck's constant, $k_\text{F}\!=\!\pi n_\text{1D}$ is the Fermi wavevector and $n_\text{1D}$ is the 1D density. An impurity of identical mass $m$ interacts with the gas through a repulsive local potential. The impurity-host interaction is characterized by $\gamma_i$. The minimal energy state of such an interacting system with finite momentum is referred to as a moving polaron state.  It represents the impurity dressed and moving together with a cloud of particles of the gas. The minimal possible energy $\varepsilon_\text{pol}(k)$ of a moving polaron state with momentum $\hbar k$ has the shape illustrated in the right panel of Fig.~\ref{fig:Fig_1}\textbf{d}. This state, recently experimentally realized~\cite{meinert2017bloch,Dhar2025}, features emergent anyonic correlations that result in asymmetric momentum distribution of the impurity for all values of $k$ except $0$ and $k_\text{F}$~\cite{Dhar2025}. In our experiment, the bare impurity injected with finite momentum $\hbar  Q$ starts away from the polaronic ground state, see middle panel of Fig.~\ref{fig:Fig_1}\textbf{d}. The subsequent temporal evolution can be viewed as a series of quantum collisions through which the impurity transfers its energy and momentum to the host until it reaches a superposition of polaron states with different momenta~\cite{SM}.

We start the experiment by adiabatically loading an optically trapped, interaction-tunable $^{133}$Cs Bose-Einstein condensate (BEC)~\cite{Kraemer2004} into an array of vertically oriented 1D tubes formed by a 2D optical lattice as illustrated in Fig.~\ref{fig:Fig_1}{\bf{a}}. The atoms are initially in the magnetic hyperfine sublevel $(F, m_\text{F})\!=\!(3,3)$ and they are levitated against gravity by means of a magnetic field gradient of $|\nabla B_z|\!\approx\!31.1$ G/cm along the vertical $z$ direction. With about $1.5\times10^5$ atoms in the initial BEC, we fill approximately 5800 tubes with an average filling of $39$ atoms per tube~\cite{SM}. A weak harmonic confinement along the tube direction, with a trap frequency of $17.0(1)$ Hz, results from the Gaussian nature of the lattice beams. By applying an offset magnetic field $B_z$ we tune the s-wave scattering length $a$ through a Feshbach resonance, see Fig.~\ref{fig:Fig_1}{\bf{c}}. Initially, the offset magnetic field is set to $20.6(1)$ G, at which $a\!=\!199(4)\,a_0$. With an average 1D density of $n_\text{1D}\!=\!1.6(1)$/$\mu$m, we have $\gamma\!=\!2.5(2)$. The average Fermi time is $t_\text{F}\!=\!166(10)~\mu$s. The impurity particles are created by transferring on average one atom per tube to the sublevel $(3,2)$ by means of a $20$-$\mu$s radio-frequency (RF) pulse~\cite{SM} (see Fig.~\ref{fig:Fig_1}\textbf{b}). The atoms in the $(3,3)$ state, constituting the majority fraction, now serve in each tube as the background host gas in which the impurity atoms in $(3,2)$ state move about. For the initial choice of magnetic field $B_z$, the impurity-host scattering length $a_i$ is zero (see Fig.~\ref{fig:Fig_1}{\bf{c}}), making the host gas transparent for the impurities with $\gamma_i\!=\!0$. Immediately after their creation, the impurity particles, given their lower magnetic moment, accelerate along the $z$ direction subject to $1/3$ of the gravitational force. When the impurities have reached the desired mean momentum $\hbar Q$, i.e. after about $0.70$ ms for $Q\!=\!k_\text{F}$, the levitation gradient is turned off, releasing the host gas and the impurities into a free fall within their respective tubes. All subsequent dynamics hence takes place in the accelerated frame of reference. Simultaneously, $\gamma_i$ is ramped to a value in the range of $\gamma_i\!=\!0$ to $12$. The ramping-time is approximately $50\,\mu$s, which is a fraction of $t_\text{F}$. The system is subsequently evolved for a certain time $t$. The maximum timescale we consider is $12$ $\,t_\text{F}$, which is much shorter than the time scale associated with the longitudinal harmonic trapping. The host and the impurity particles are imaged simultaneously after release into 3D and $20$-ms levitated time-of-flight (ToF). The impurity-host scattering length $a_i$ is switched back to zero for a faithful measurement of the impurities' momentum distribution. The presence of the levitation field during ToF spatially separates the host and impurity particles along the $z$ direction. During the levitated expansion, the host-gas interactions, $a\approx 199\,a_0$, are finite.

\begin{figure}[t!]
\centering
\renewcommand{\figurename}{FIG.}
\includegraphics[trim={0cm 0cm 0cm 0cm},clip,width=1\columnwidth]{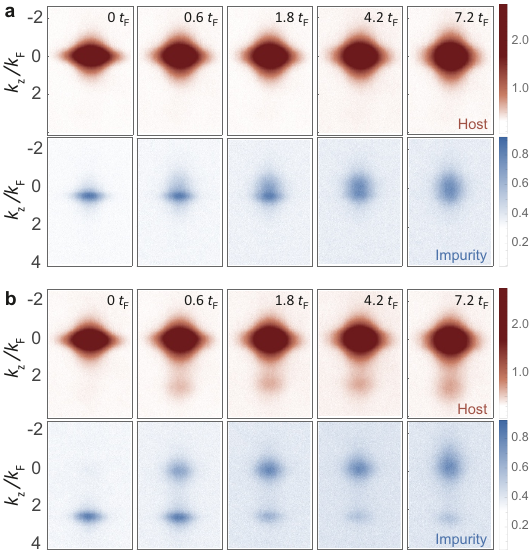}
	\caption{\textbf{Subsonic versus supersonic impurity dynamics.} 
   Absorption images for initially \textbf{a}, subsonic ($Q\!=\!0.40(1)\, k_\text{F}$) and \textbf{b}, supersonic impurities ($Q\!=\!2.30(8)\, k_\text{F}$), showing the host (top panels, red) and impurity (bottom panels, blue) distributions during the relaxation process for the times indicated. The interactions are set to $(\gamma_i,\gamma)\!=\!(2.6(3),5.9(3))$. Each image is an average of 10 experimental realizations. The vertical axis in each image is converted to a momentum scale. The horizontal axis is not calibrated as this direction is perpendicular to the tubes and no relevant dynamics occur along this direction.}
	\label{fig:Fig_2}
\end{figure}

Our analysis of the relaxation process is based on absorption images of the host and impurity atoms as shown in Fig.~\ref{fig:Fig_2}. For the impurities, these reflect their momentum distribution, while this is only approximately true for the host atoms due to the finite but non-zero interactions during ToF. The initial impurity distributions have rms widths of about $0.3 \,\hbar k_\text{F}$ in the $k_z$-direction. For the subsonic case in Fig.~\ref{fig:Fig_2}\textbf{a}, both the impurity and the host distributions broaden but remain singly peaked. This behavior is strikingly different for the supersonic case, shown in Fig.~\ref{fig:Fig_2}\textbf{b}. There, within about one Fermi time, separate momentum peaks appear, one at $k_z/k_\text{F}\!\approx\!0$ for the impurities and one centered around the initial momentum for the host. Both of these peaks have the same population within our measurement accuracy~\cite{SM}. 
The emergence of such a peak in the host indicates the formation of a shock wave in the quantum fluid triggered by the supersonic impurity.

\begin{figure*}[t!]
\centering
\renewcommand{\figurename}{FIG.}
\includegraphics[trim={0cm 0cm 0cm 0cm},clip,width=\textwidth]{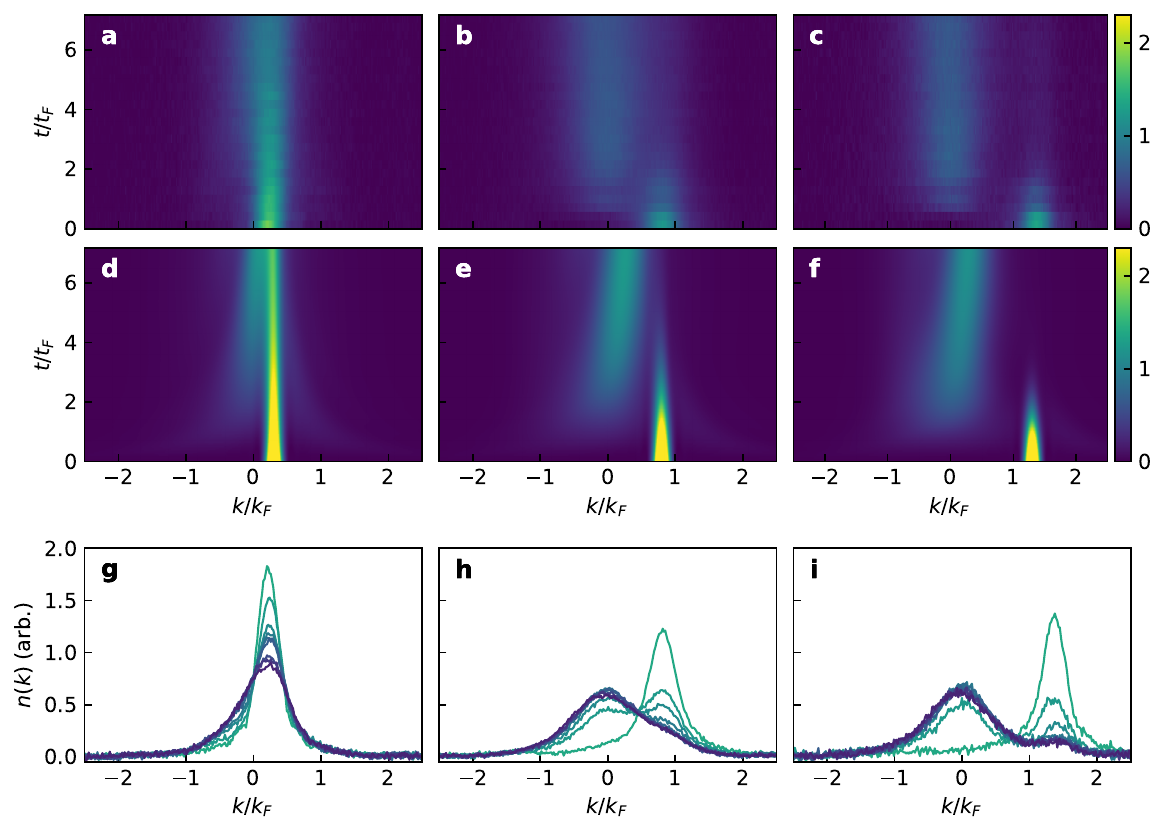}
	\caption{{\bf Time evolution of the impurity momentum distribution $n(k)$.} {\bf{a, b, c}}, Experimental $n(k)$ for $(\gamma_i,\gamma)\!=\!(11.0(7),18.5(12))$ and selected values of the initial impurity wavevector $Q/k_\text{F} \!=\! 0.2, 0.9, 1.4$, respectively, for times up to $t = 7$ $t_\text{F}$ in time steps of $0.6$ $t_\text{F}$. {\bf{d, e, f}}, Numerical results for $(\gamma_i,\gamma)\!=\!(9.9, 18.7)$ and $Q/k_\text{F} \!=\! 0.3, 0.8, 1.3$, respectively, as discussed in the text. {\bf{g, h, i}}, Experimental distributions for from $t\!=\!0$ to $t\!=\!7 t_\text{F}$ in steps of $0.9 t_\text{F}$ (from green to purple). Each experimental dataset is the average of $10$ realizations.}
	\label{fig:Fig_3}
\end{figure*}

For a more detailed analysis, we integrate the data along the transverse direction and obtain momentum distributions $n(k)$ for the impurity atoms along the $(k\!=\!k_z)$-axis, shown in Fig.~\ref{fig:Fig_3}\textbf{a}, \textbf{b}, and \textbf{c} for $Q/k_\text{F}\!=\!0.20(1), 0.90(2)$, and $1.40(6)$, respectively. Selected distributions are presented in Fig.~\ref{fig:Fig_3}\textbf{g}, \textbf{h}, and \textbf{i}. In each case, the distribution broadens and relaxes as the population is transferred to lower momenta. Specifically, for the subsonic case with $Q/k_\text{F}\!=\!0.20(1)$ the distribution develops a shoulder on the low-momentum side. 
While the mean momentum changes to $0.09(2)\,\hbar k_\text{F}$, the most probable momentum remains near $0.20(1)\,\hbar k_\text{F}$. In the intermediate regime with $Q/k_\text{F}\!=\!0.90(4)$, the distribution relaxes to a comparatively broad distribution with a mean at $0.06(2)\,\hbar k_\text{F}$. For the supersonic case $Q/k_\text{F}\!=\!1.40(6)$ the initial and final distributions are largely separated, with a transient doubly-peaked distribution. The contribution at the initial momentum rapidly decays on the time scale of $t\!\approx\!t_\text{F}$.

Our experimental findings are  supported by numerical simulations based on matrix product states~\cite{Hauschild2018, Knap2014Flutter}. These simulations model a single 1D system governed by the many-body Hamiltonian
\begin{equation}
    \hat{H} = \hat{H}_\text{LL}(g_\text{1D}, \{z_n\}) + \frac{\hat{P}_\text{imp}^2}{2m}+g_i\sum_{n=1}^N \delta(z_{n}-z)
\label{eq:1}
\end{equation}
where, the Lieb-Liniger Hamiltonian $\hat{H}_\text{LL}$ describes a gas of $N$ bosons with coordinates $z_n$, interacting via repulsive contact interactions of strength $g_\text{1D}\!=\!\hbar^2\gamma  n_\text{1D}/ m$. The second term corresponds to the kinetic energy of the impurity, with momentum $P_\text{imp}$, while the last term describes the interaction between the impurity and the host particles with interaction strength $g_i\!=\!\hbar^2\gamma_i n_\text{1D} / m$. In the simulations, at $t\!=\!0$ the impurity is injected as a wave packet with a mean momentum $Q$ into the ground state of the host gas. Similarly to the experiment, $\gamma_i$ and $\gamma$ are simultaneously quenched from $0$ to $9.9$ and from $4.5$ to $18.7$, respectively. The subsequent time evolution of the impurity momentum distribution presented in Fig.~\ref{fig:Fig_3}\textbf{d} to \textbf{f} agrees well with the experimental data in Fig.~\ref{fig:Fig_3}\textbf{a} to \textbf{c}. We observe minor quantitative differences, specifically for the case of large $Q$. In the experiment, after relaxation a systematically larger impurity fraction propagates freely through the host gas compared to the theoretical predictions. We attribute this to contributions from outer tubes that contain fewer particles. At large $Q$, the impurities are accelerated out of these shorter tubes before the host-impurity interaction is switched on~\cite{SM}. One further difference between the experiment and the simulations lies in the injection protocol of the impurity: the impurity is created out of the host via the aforementioned RF pulse in the former, and it is injected into the sample as a wave-packet in the latter.

The far-from-equilibrium initial impurity evolves to a steady state that can be interpreted as a superposition of finite-momentum polaron states~\cite{Gamayun2020}. Each of these polaron states for $0<Q<k_\text{F}$ exhibits an asymmetric momentum distribution for the impurity. This is reflected in the asymmetric shape of our relaxed momentum distribution for small $Q$, see Fig.~\ref{fig:Fig_3}\textbf{g}. In the opposite case, $Q>k_\text{F}$, the shape of the relaxed momentum distribution, as seen in Fig.~\ref{fig:Fig_3}\textbf{i}, is rather distinct from the polaronic distribution around the same $Q$~\cite{SM}. Instead, it resembles the momentum distribution around $k_\text{F}$. This relaxation picture is consistent with the one given in Refs.~\cite{mathy2012quantum,Knap2014Flutter} due to the presence of a large density of states near $k_\text{F}$~\cite{Dolgirev2021}. During relaxation, long-lived coherent oscillations for $Q>k_\text{F}$, termed quantum flutter, were predicted. Observing these relatively low-magnitude oscillations would require further improvements of our experiment.

\begin{figure}[t!]
\centering\
\renewcommand{\figurename}{FIG.}
\includegraphics[trim={0cm 0cm 0cm 0cm},clip,width=\columnwidth]{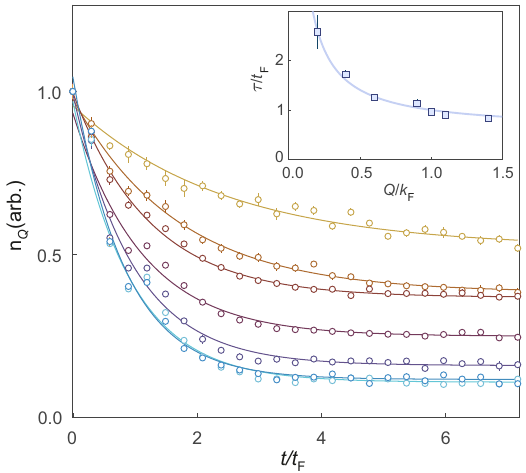}
	\caption{{\bf{Ultrafast relaxation dynamics of the impurity}}. The time evolution of $n_Q$ is measured for $(\gamma_i,\gamma)\!=\!(11.0(7),18.5(12))$ and distinct momenta that, from slowest to fastest decay (yellow to blue), are $Q/k_\text{F}\!=\! 0.2, 0.4, 0.6, 0.9, 1.0, 1.1, 1.4$. The solid curves correspond to exponential fits including an offset. Each data point is the average of 10 repetitions. The error is the standard error of the mean. Inset: Time constant $\tau$ extracted from the exponential fits as a function of $Q$. The solid curve is a phenomenological fit with the function $a(k_\text{F}/Q) + c$, where $a\!=\!0.41(3) $, and $c\!=\!0.64(4) $. } 
	\label{fig:Fig_4}
\end{figure}

Our data in Fig.~\ref{fig:Fig_3} shows that relaxation is faster for a higher initial momentum $\hbar Q$. To quantify this, we track the evolution of the momentum population $n_Q \equiv n(k=Q)$ at the location $Q$ of the peak of the initial distribution, averaged over $\pm 0.1\,k_\text{F}$ around $Q$. The results presented in Fig.~\ref{fig:Fig_4} for different values of $Q$ demonstrate that the initial peak depletes faster for higher $Q$. Around $Q\!\approx\!k_\text{F}$ the relaxation time settles to values near $t_\text{F}$, see inset of Fig.~\ref{fig:Fig_4}.

\begin{figure}[b!]
\renewcommand{\figurename}{FIG.}
\centering
\includegraphics[trim={0cm 0cm 0cm 0cm},clip,width=\columnwidth]{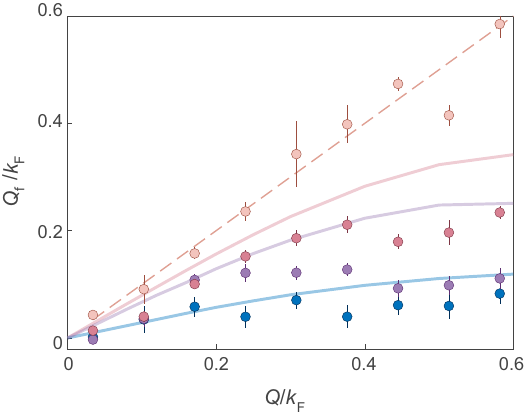}
\caption{ \textbf{Dissipationless motion of the impurity after relaxation.} The mean steady-state momentum of the impurity $Q_\text{f}/k_\text{F}$ as a function of its initial momentum $Q/k_\text{F}$, for different values of interaction strength $\gamma_i$. Pink, red, purple and blue datasets correspond to $(\gamma_i, \gamma)\!=\! (0, 2.6 )$, $(1.9,5.1)$, $(3.5,7.4)$, and $ (11.5, 19.4)$, respectively. Each data point is an average of 10 repetitions except for the non-interacting case ($\gamma_i\!=\!0$), where 3 repetitions were taken. The error is the standard error of the mean. 
The dashed line is the predicted momentum of a non-interacting particle. The solid curves are results from numerical simulations.
}
	\label{fig:Fig_5}
\end{figure}

We now examine the steady-state momentum of the impurity, $\hbar Q_\mathrm{f}$, in the long-time limit. The final momentum is extracted from the impurity momentum distribution averaged over the interval 
$[-2\hbar k_\mathrm{F},\,2\hbar k_\mathrm{F}]$ after an evolution time $t\!=\!12\,t_\mathrm{F}$. Figure~\ref{fig:Fig_5} shows $Q_\mathrm{f}$ as a function of the initial impurity wavevector $Q$ for four different interaction strengths $\gamma_i$. In the absence of interactions ($\gamma_i\!=\!0$), the final momentum follows the trivial linear dependence expected for a freely propagating impurity. For finite $\gamma_i$, the impurity reaches a nonzero steady-state momentum. At small initial wavevectors, $Q_\mathrm{f}$ increases with $Q$, and gradually approaches a constant value at larger $Q$. The slope of $Q_\mathrm{f}$ at low $Q$ decreases as $\gamma_i$ increases, and the saturation value of $Q_\mathrm{f}$ is reduced for stronger interactions. These observations of non-zero $Q_\mathrm{f}$ of the impurity in the steady-state are supported by our numerical simulations. For weak interactions, the simulated values of $Q_\mathrm{f}$ lie above the corresponding experimental results, which we attribute to differences in the experimental and numerical protocols discussed above.

Although our initial state realizes a finite energy state above the polaron branch, we can gain some intuition of why $Q_\text{f}$ is non-zero by utilizing the polaron picture~\cite{SM}. The group velocity of the polaron is
\begin{equation}
v_\text{pol} =  \frac{\partial \varepsilon_\text{pol}(k)}{\partial k}.
\end{equation}
The average velocity of the impurity particle in the polaron state is equal to $v_\text{pol}$~\cite{Knap2014Flutter}. The shape of $\varepsilon_\text{pol}(k)$, illustrated in Fig.~\ref{fig:Fig_1}\textbf{d} for a finite impurity-gas repulsion, is quadratic at small $k$, $\varepsilon_\text{pol}(k) =(\hbar k)^2/2m_\text{pol}$, consistent with the common understanding of a small-momentum polaron as a bare impurity with a renormalized mass $m_\text{pol} \ge m$. The observation that $v_\text{pol}$ is a linear function of $k$ for a small-momentum polaron is consistent with the linear dependence of $Q_\text{f}$ on $Q$. This is also reflected in our data, Fig.~\ref{fig:Fig_5}. A deviation from linearity in our data for higher values of $Q$ indicates the breakdown of the quadratic regime for $\varepsilon_\text{pol}(k)$. The decrease in $Q_\text{f}$ with increasing $\gamma_i$ is also consistent with flattening out $\varepsilon_\text{pol}(k)$. 
For larger $\gamma_i$, an increasing number of excitations from the surrounding medium dresses the impurity. 
This results in an increase of $m_\text{pol} \ge m$ of the polaron. 
Beyond the effective polaron picture, the flattening of the dispersion can also be seen in the exactly solvable limit of infinitely strong background interaction, which is described by McGuire's model~\cite{SM}. Crucially, though, the existence of a polaron branch, which enables the impurity's dissipationless motion, is a generic feature of quantum dynamics in 1D, and is not tied to a particular form of interparticle interactions.

In contrast to the Landau criterion for macroscopic obstacles in superfluids, we have experimentally demonstrated that a microscopic quantum impurity does not come to a full stop when injected into a one-dimensional quantum fluid. Instead, the impurity evolves into a correlated many-body state that propagates at a finite velocity. For initially supersonic impurities, this relaxation is accompanied by the emission of shock waves and occurs on the timescale of a Fermi time.

Our results pave the way for a wide range of future studies. Of particular interest is the prospect of observing the formation of bipolarons~\cite{bipolaron2018,will2021bipolaron} by increasing the impurity density in our system. Realizing a more homogeneous system in a box-shaped trap would likely allow the observation of coherent oscillations of the impurity referred to as quantum flutter~\cite{mathy2012quantum, Knap2014Flutter}. Altering the transverse confinement strength allows for exploring dimensional crossover effects~\cite{Guo2024}, including emergent quantum Cherenkov radiation~\cite{Seetharam2021cerenkov,Seetharam2024cerenkov}. Our experiment demonstrates how the interplay of quantum effects can give rise to surprising transport phenomena, 
opening new strategies for the controlled propagation of information and suppression of dissipation in quantum many-body systems.

\bigskip

\noindent{\bf Acknowledgments}\\
The Innsbruck team acknowledges funding by a Wittgenstein prize grant under Austrian Science Fund (FWF) project number Z336-N36, by the European Research Council (ERC) under project numbers 789017 and 10120161, by an FFG infrastructure grant with project number FO999896041. The research was funded in part by the FWF 10.55776/COE1 and the European Union – NextGenerationEU. MH thanks the doctoral school ALM for hospitality, with funding from the FWF under the project number W1259-N27. Y.G. is supported by the FWF with project number 10.55776/COE1. MK acknowledges support from the Deutsche Forschungsgemeinschaft (DFG, German Research Foundation) under Germany’s Excellence Strategy--EXC--2111--390814868, TRR 360 – 492547816 and DFG grants No. KN1254/1-2, KN1254/2-1, the European Union (grant agreement No. 101169765), as well as the Munich Quantum Valley (MQV), which is supported by the Bavarian state government with funds from the Hightech Agenda Bayern Plus. 
\textbf{Author Contributions:} The work was conceived by H.C.N., M.B.Z., M.K., M.H., S.D., M.L. Experiments were prepared and performed by M.H. and S.D. Data were analyzed by M.H. and S.D. Numerical simulations were performed by E.W. and M.K. The manuscript was drafted mainly by S.D., M.H., M.B.Z., M.K., M.L., and H.C.N. All authors contributed to the discussion and finalization of the manuscript.
{\bf Data Availability:} The data shown in the main text are available via Zenodo~\cite{Zenodo}.
{\bf Code Availability:} Codes supporting the findings of this study are available from the corresponding author on a reasonable request.

\newpage
\bibliography{References.bib}

\clearpage
\onecolumngrid
\newpage

\setcounter{equation}{0}  
\setcounter{figure}{0}
\setcounter{page}{1}
\setcounter{section}{0}    
\renewcommand\thesection{\arabic{section}}    
\renewcommand\thesubsection{\arabic{subsection}}    
\renewcommand{\thetable}{S\arabic{table}}
\renewcommand{\theequation}{S\arabic{equation}}
\renewcommand{\thefigure}{S\arabic{figure}}
\setcounter{secnumdepth}{2}  

\begin{center}
{\large \textbf{Supplementary Materials of\\ ``\titleinfo''}}\\
\vspace{7pt}

Milena Horvath, Sudipta Dhar, Elisabeth Wybo, Dimitrios Trypogeorgos Yanliang  Guo,  Mikhail Zvonarev, \\ Michael Knap, Manuele  Landini, Hanns-Christoph  N{\"a}gerl
\vspace{5pt}

\end{center}
\twocolumngrid

\section{Preparation of 1D tubes}
\label{sec:Exp_prep}
The experiment starts with a 3D interaction-tunable BEC consisting of approximately $1.5\!\times\!10^5$ $^{133}$Cs atoms in the absolute ground state $(F, m_\text{F})\!=\!(3,3)$, confined in a crossed-beam dipole trap~\cite{Kraemer2004} with trapping frequencies of $(\omega_x, \omega_y, \omega_z)/2\pi \!=\!(5.7(1), 11.8(1), 10.3(1))$ Hz, giving Thomas-Fermi radii of $(21.4(2), 16.0(1), 16.9(1))\,\mu$m. The BEC is adiabatically loaded into an array of 1D tubes created by a 2D optical lattice. The laser beams forming the lattice propagate in the horizontal $x$-$y$ plane at right angle to each other, and their power is ramped up in 500 ms to give a lattice depth of $30 E_\mathrm{r}$, where $ E_\mathrm{r} = \pi^{2}\hbar^{2}/2md^{2}$ is the recoil energy. Here, $d\!=\!\lambda/2\!=$532 nm is the lattice spacing set by the wavelength $\lambda$ of the lattice light. At this depth, tunneling between the tubes is fully suppressed on the considered timescales, and the tubes can be considered independent. Due to the underlying dipole trap during the lattice loading, the resulting array of tubes has a non-homogeneous distribution for the number of atoms. We have on average 39 atoms per tube. Additionally, the Gaussian profile of the lattice beams results in weak harmonic trapping with trap frequency $\omega_{z}/2\pi=17.0(1)$ Hz along the tubes' direction.

\section{Experimental system parameters}
\label{sec:Exp_params}
By controlling the scattering length $a$ we set the 1D coupling constant $g_\text{1D}$~\cite{Haller2009Realization, OlshaniiAtomic1998} via 
\begin{equation}
    g_\text{1D} = 2\hbar \omega_\perp a \left ( 1-1.0326 \frac{a}{a_\perp} \right ),
\end{equation}
where $a_\perp = \sqrt{\hbar/m\omega_\perp}$ is the transversal oscillator length. Similarly, the scattering length $a_i$ for collisions between the impurity and the host atoms sets the coupling constant $g_i$. The density of the host gas $n_{1\text{D}}$ then determines the dimensionless interaction strengths $\gamma$ and $\gamma_i$. This requires an estimate for the atom number $N_{j,k}$ for each tube $(j, k)$. Assuming that the interactions are sufficiently small during the adiabatic loading of the lattice, such that all tubes are in the Thomas-Fermi regime, the atom number $N_{j,k}$ for the tube can be computed iteratively from the global chemical potential. For further details, see~\cite{MeinertProbing2015}. We then calculate the average number of atoms per tube $N\!=\!\frac{1}{N_{tot}}\sum_{j,k} N_{j,k}^2$, where $N_\text{tot}\!=\!\sum_{j,k} N_{j,k}$ is the total number of atoms. The 1D density distribution $n(z)$ depends on the interaction regime in which the gas is in. Here, we calculate $n(z)$ by exactly solving the Lieb-Liniger model within the local density approximation (LDA)~\cite{LiebExact1963, DunjkoBosons2001}. We then calculate the average 1D density $n_\text{1D}=\frac{1}{N} \int n(z)^2 dz$. The average Fermi wavevector is given by $k_\text{F}=\pi n_\text{1D} $. The distribution of tube lengths causes the value of $k_\text{F}$ to vary from tube to tube. The average interaction strengths are calculated as
\begin{equation}
    \gamma= \frac{mg_\text{1D}}{\hbar^2n_{1 \text{D}}} \quad\quad \text{and} \quad\quad
    \gamma_i = \frac{mg_\text{i}}{\hbar^2n_{1 \text{D}}}.
\end{equation}

\begin{figure}[b!]
    \centering
\includegraphics[trim={0cm 0cm 0cm 0cm},clip,width=\columnwidth]{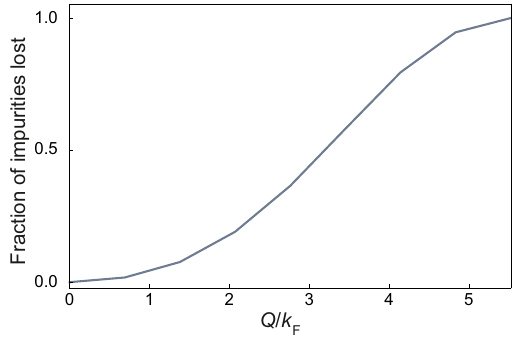}
\caption{ \textbf{Impurities leaving the host gas.} Estimated fraction of impurities that have left the host gas and hence are lost to the system as a function of initial $Q/k_\text{F}$. Note that the experiments reported in the main text are done for initial momenta below $2.3\,k_\text{F}$, and in particular the experiments on the dissipationless flow have been carried out for $Q/k_\text{F}\leq0.6$. }
	\label{fig:Sup_mat_leftimp}
\end{figure}

\begin{figure*}[t!]
\centering
\includegraphics[trim={0cm 0cm 0cm 0cm},clip,width=\textwidth]{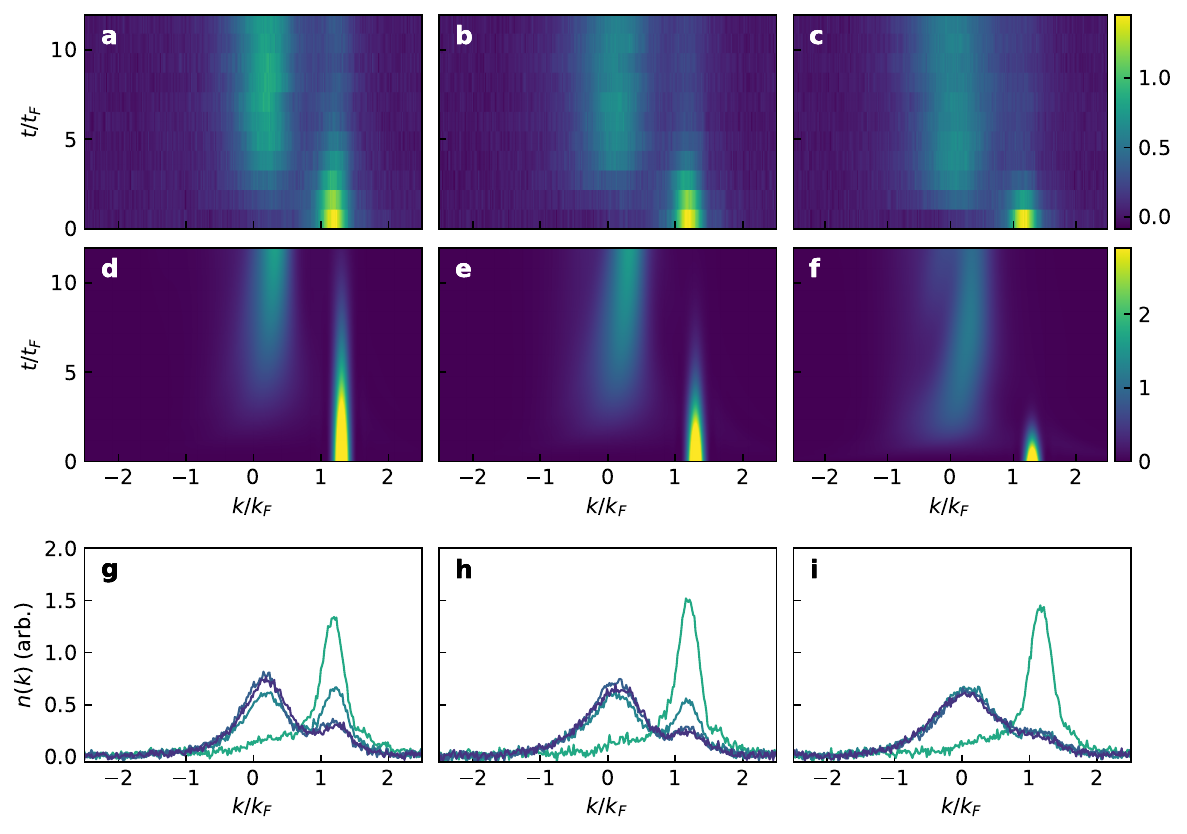}
\caption{ \textbf{Momentum distributions for different values of the host-impurity interaction strength $\gamma_i$.} Experimental data (top row) and theory (middle row). {\textbf{a, b, c}}, Experimental $n(k)$ at a fixed $Q \!=\! 1.1\,k_\text{F}$ for $\gamma_i\!=\! 2.9(1), 4.1(2)$, and $9.1(4)$, respectively. {\textbf{d, e, f}},  numerical results at a fixed $Q \!=\! 1.3 \,k_\text{F}$ for $\gamma_i \!=\! 2.9, 4.5$, and $9.9$, respectively. {\bf{g, i, j}}, Experimental $n(k)$ for some specific evolution times between $t\!=\!0$ to $12\,t_\text{F}$ (from green to blue).}
	\label{fig:Sup_mat_interaction_dynamics}
\end{figure*}

\begin{figure*}[t!]
\centering
\includegraphics[trim={0cm 0cm 0cm 0cm},clip,width=\textwidth]{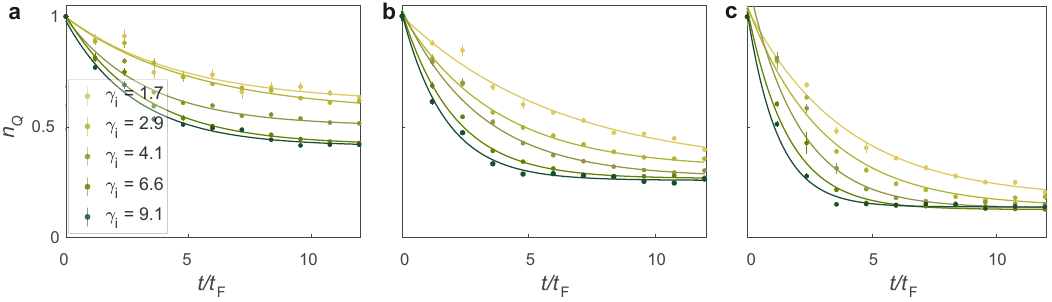}
\caption{ \textbf{Evolution of $n_Q$ for different values of $\gamma_i$.}  { \bf{a, b, c}}, $n_Q$ as a function of time for $Q/k_\text{F} \!=\! 0.2,\, 0.6,$ and $1.1$, respectively. For each $Q$, the decay of $n_Q$ is measured for four different values of $\gamma_i$ indicated in the first panel. The solid curves are obtained by fitting the data by an exponential function with an offset. Each data point is an average of 5 to 10 repetitions.}
	\label{fig:Sup_mat_Interaction_dependence}
\end{figure*}

\begin{figure}[t!]
\centering
\includegraphics[trim={0cm 0cm 0cm 0cm},clip,width=\columnwidth]{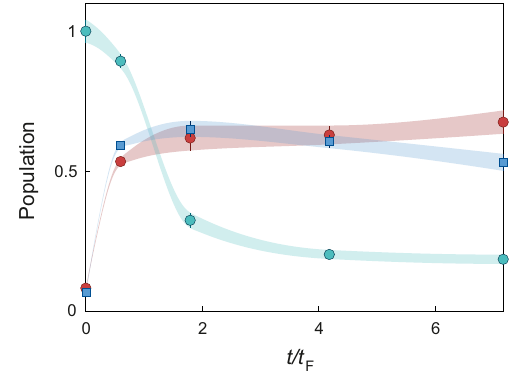}
\caption{ \textbf{Supersonic impurity dynamics.} Normalized population in the $n(k\!=\!Q)$-bin of the host distribution (red circles) and of the impurity distribution (green circles). The blue squares show the population in the impurity $n(k\!=\! 0)$ bin. For this dataset $Q \!=\! 2.30(8)$ $k_\text{F}$ and $(\gamma_i, \gamma) \!=\! (2.6(3), 5.9(3))$. Each data point is the average of 10 repetitions. The error bars reflect the standard error and the shadings are a guide to the eye.}
	\label{fig:Sup_mat_Ejected_particle_population}
\end{figure}

\begin{figure*}[t!]
\centering
\includegraphics[trim={0cm 0cm 0cm 0cm},clip,width=\textwidth]{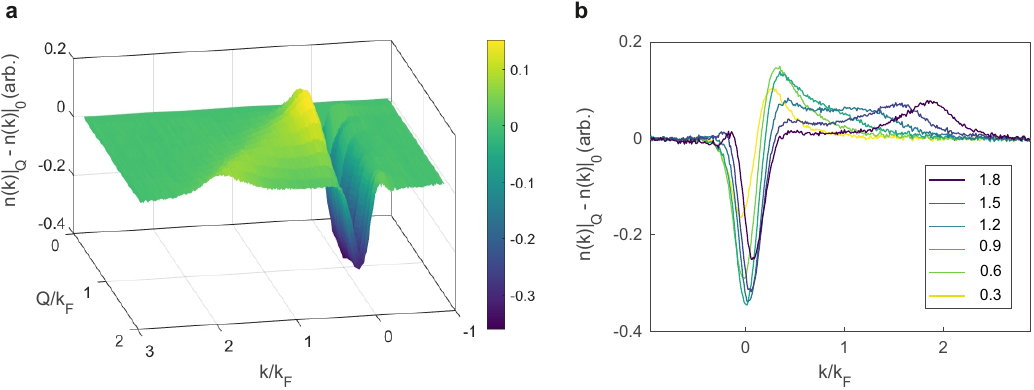}
\caption{ \textbf{Measured wavepacket emission of host particles.} \textbf{a}, \textbf{b}, Change in the momentum distribution of the host gas $n(k)|_{Q} - n(k)|_{Q=0}$ as a function of $Q$. \textbf{b}, Distributions for selected values of $Q/k_\text{F}$. The corresponding values of $Q/k_\text{F}$ are given in the legend. Each distribution is the average of ten repeats. 
}	\label{fig:Sup_mat_Ejected_host_exp}
\end{figure*}
\section{Impurity creation via RF transfer}
The vertically oriented magnetic gradient field $\nabla B_{z}$ gives rise to a position-dependent resonance frequency for the transfer of the host atoms to the $(3,2)$ state. The creation of the impurity can thus be made spatially selective along the tube. A $20$-$\mu$s RF pulse is used to create the impurity. From the length of the RF pulse, we estimate the spread of the impurity wavepacket to be $\sigma_z\! \sim\! 16\,\mu$m. This corresponds to $\sigma_\text{k}\!\sim\! 0.02\,k_\text{F}$ in the momentum space. Note that the measured width of the impurity in the momentum space at $t\!=\!0\,t_\text{F}$ in Fig.~\ref{fig:Fig_2} in the main text is $\sim 0.3k_\text{F}$. We attribute this broadening to the inhomogeneous tube distribution, finite ToF, and to the finite coherence length of the host gas out of which the impurity is created. After impurity creation, during the acceleration phase, impurities in shorter tubes may be driven beyond the spatial extent of the host atoms before the impurity-host interaction is switched on. We simulate the fraction of impurities that leave the host gas based on the knowledge of the atomic density distribution across the tubes, duration of RF pulse, and the magnetic field gradient. The results are shown in Fig.~\ref{fig:Sup_mat_leftimp}. For $Q\approx k_{F}$ this fraction is below $10\%$. For $Q>5\,k_\text{F}$ nearly all impurities would have left the host gas in all tubes.

\section{Interaction dependence}

In the main text, the discussion focused on the strongly interacting regime with $\gamma_i\!=\!11.0(7) \gg 1$, where the bosons are expected to be fermionized. Figure~\ref{fig:Sup_mat_interaction_dynamics} presents both experimental and simulated results for an impurity with $Q \!=\! 1.1\,k_\text{F}$ for three different values of the interaction strength: $\gamma_i\!=\!2.9(1)$, $4.1(2)$, and $9.1(4)$. We observe that stronger interactions lead to faster relaxation dynamics. We find good qualitative agreement between experiment and simulation. We further quantify the interaction dependence using the $n_Q$-momentum bin. Figures~\ref{fig:Sup_mat_Interaction_dependence}~\textbf{a}-\textbf{c} show the dependence of $n_Q$ on the interaction strength for three representative regimes: $Q \ll k_{\text{F}}$, $Q \approx k_{\text{F}}$, and $Q > k_{\text{F}}$. In all three regimes, the decay time of $n_Q$ decreases with increasing interaction strength, and the steady-state value of $n_Q$ decreases with increasing $Q$.

\section{Ejected host particle}

For sufficiently large $Q$, we observe the emission of a wave packet from the host momentum distribution (Fig.~\ref{fig:Fig_2}\textbf{b}). Figure~\ref{fig:Sup_mat_Ejected_particle_population} shows the normalized population within a momentum bin of width $0.3\,k_{\mathrm{F}}$, centered at $k \!=\! Q$, for both the impurity and host distributions, as well as at $k\!=\!0$ for the impurity, for the data shown in Fig.~\ref{fig:Fig_2}\textbf{b}. Within $t \!=\! 2\,t_{\mathrm{F}}$, the impurity population at $k \!=\! Q$ decreases by approximately $80\%$, while the populations at $k \!=\! Q$ in the host and at $k \!=\! 0$ for the impurity increase to approximately $70\%$. This behavior confirms the momentum transfer from the mobile impurity to the host gas.
 
\section{Momentum distribution of the host gas}
The injection of a mobile impurity into a strongly interacting Bose gas can give rise to a shock wave in the form of an ejected particle from the host distribution. Here, we measure the momentum distribution of the host gas for different values of $Q/k_\text{F}$. For a faithful measurement of the momentum distribution of the host, here we set the host-gas scattering length to $a\approx0\,a_0$ during ToF. In our experiment, quenching the impurity-host interaction is accompanied by a quench of the host-host interactions, leading to a broadening of the host momentum distribution during the subsequent evolution. To isolate momentum transfer from the impurity to the host, we fix the evolution time and subtract the host momentum distribution measured for a stationary impurity ($Q/k_\text{F}=0$) from that obtained for a finite $Q$. The results are presented in Fig.~\ref{fig:Sup_mat_Ejected_host_exp}. We find that for $Q\simeq k_\text{F}$ an ejected wavepacket develops that carries away most of the initial momenta of the impurity.

\section{Theory of Superfluidity}

In this Section, we elaborate on how collisions of an impurity with the surrounding gas are determined by kinematic (that is, stemming from energy and momentum conservation laws) restrictions. They come from an interplay of quantum statistics and interaction effects constrained by the 1D geometry of our problem. 
\begin{figure}[b!]
\centering
\includegraphics[trim={0cm 0cm 0cm 0cm},clip,width=\columnwidth]{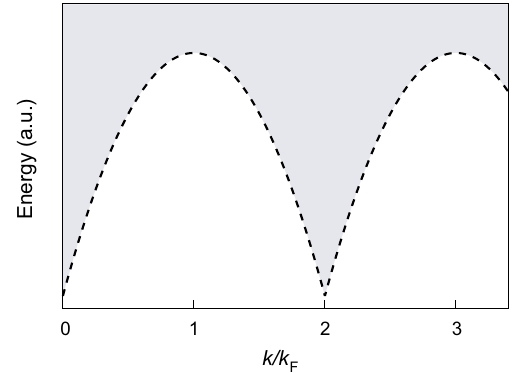}
\caption{\textbf{Excitation spectrum of a 1D gas}. Excitation energy as a function of total momentum. The shaded region indicates the allowed excitations continuum. Dashed black line is the exact lower bound (edge) of the excitation spectrum.}
	\label{fig:multispectrum}
\end{figure}

We illustrate in Fig.~\ref{fig:multispectrum} the excitation spectrum of indistinguishable quantum particles constrained to one spatial dimension. Several crucial features remain in this picture regardless of particle statistics and the strength of interparticle repulsion~\cite{giamarchi_book_1d}. The first is the presence of the exact nonzero lower bound (or edge), $\varepsilon_\text{min}(k)$, of the particle-hole continuum, shown by a dashed curve in Fig.~\ref{fig:multispectrum}, as well as in Fig.~\ref{fig:Fig_1} of the main text. 
The momentum $p$ is related to the wave number $k$ as $p=\hbar k$. All possible excitations of the quantum gas stay above this curve. The second is the fact that the edge energy at zero momentum has the same value as at $k=2k_\text{F}$, independent of the interaction strength and quantum statistics of the constituent particles (this is referred to as the Luttinger theorem). The third is the $2k_\text{F}$-periodicity of the edge (in the thermodynamic limit). This periodicity is the reason for the frequent usage of the term ``first Brillouin zone'' for the interval $0<k<2k_\text{F}$ even in \textit{continuous} quantum systems in 1D. The fourth is the particular shape of the edge near $k=0$: it has a linear slope:
\begin{equation}
\varepsilon_\text{min}(k) = \hbar vk, \qquad k\to0. \label{lindisp}
\end{equation}
The low energy and momentum excitations in such a system are, therefore, sound waves whose velocity are 
\begin{equation} \label{vLLsound}
v = \left. \frac{\partial \varepsilon_\mathrm{min}(k)}{\partial \hbar k} \right|_{k=0}.
\end{equation}
Because these excitations propagate through the system by causing density fluctuations (or charge fluctuations in the case of charged particles), they are often referred to as plasmons in the literature.

\begin{figure*}[t!]
\includegraphics[trim={0cm 0cm 0cm 0cm},clip,width=\textwidth]{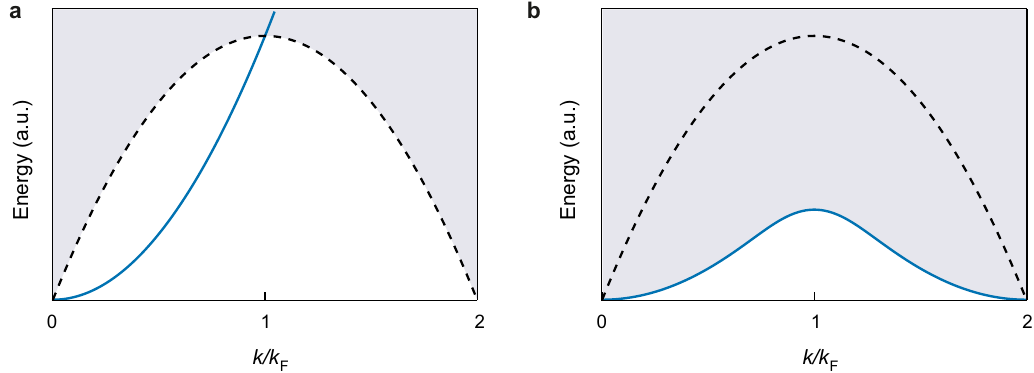}
\caption{\textbf{Emergence of the polaron spectrum due to the impurity-gas interaction}. \textbf{a}, Free impurity in the 1D gas. Particle-hole excitations of the 1D gas lie in the shaded area, above the dashed black line. The dispersion of the free impurity, given by Eq.~\eqref{freemagnondisp}, is shown by solid blue line. \textbf{b}, The impurity interacts (repulsively) with the 1D gas forming a polaron branch (blue).}
\label{fig:disprsionmp}
\end{figure*}

Next, we review in brief the arguments why the 1D system, having the excitation spectrum shown in Fig.~\ref{fig:multispectrum}, cannot be superfluid according to the Landau criterion, that we briefly recall here. Assume that the liquid moves through a pipe with velocity $\mathbf{v}$ uniformly and steadily at a given time. Landau argues that a change in the macroscopic state (velocity $\mathbf{v}$ in our case) is caused by the emergence of elementary excitations in the liquid (however, a single elementary excitation causes only infinitesimal change in the macroscopic state). Consider a single elementary excitation with momentum $\mathbf{p}$ and energy $\varepsilon(p)$ emerging in the reference frame of the liquid (that is, where the liquid is initially at rest). Here, $p$ is the length of the vector $\mathbf{p}$. Now, let us turn to the laboratory frame. There, the momentum of the liquid is
\begin{equation}
\mathbf{P}_\textrm{lab} = \mathbf{p}+M \mathbf{v},
\end{equation}
and the energy is 
\begin{equation}
E_\textrm{lab} = \varepsilon + \mathbf{p}\mathbf{v} +\frac{Mv^2}2.
\end{equation}
The term $\frac{Mv^2}2$ is the original kinetic energy of the flowing liquid of mass $M$, and $\varepsilon + \mathbf{p}\mathbf{v}$ is the change in energy due to the appearance of the excitation. Elementary excitations can only emerge when
\begin{equation}
\varepsilon + \mathbf{p} \mathbf{v} \le 0. \label{econd}
\end{equation}
 Given $p$, the minimum $v$ to satisfy the inequality~\eqref{econd} is reached when $\mathbf{p}$ and $\mathbf{v}$ are antiparallel, $\varepsilon - pv \le 0$, that is,
\begin{equation}
v \ge \frac{\varepsilon(p)}p \label{econd2} \qquad \textrm{Landau criterion}
\end{equation}
at least for some value of $p$. The inequality~\eqref{econd2} is the Landau criterion of superfluidity. When inequality~\eqref{econd2} is satisfied for some value of $p$, energy and momentum conservation \textit{do not prohibit} the emergence of excitations with energy $\varepsilon(p)$ at the expense of kinetic energy, and the liquid is not superfluid. This is always the case for 1D systems, whose excitation spectrum, illustrated in Fig.~\ref{fig:multispectrum}, contains excitations with vanishing energy while having a momentum of  $2k_\text{F}$.

Not being a superfluid means that a 1D quantum gas dissipates energy and momentum in collisions with \textit{macroscopic} obstacles. However, the motion of a \textit{microscopic} impurity can be dissipationless. To illustrate this distinction, we first consider a product state of a microscopic impurity and a background gas. In this case, momentum relaxation proceeds via the emission of quasiparticles into the fluid, accompanied by impurity recoil~\cite{Lychkovskiy2015}. We consider an impurity with initial momentum $Q$ moving through a fluid initially at rest. Before the creation of an excitation, the total momentum is $Q$ and the total energy is $E_i=Q^2/(2m)+E_0$, where $m$ is the (effective) mass of the impurity and $E_0$ is the ground state energy of the fluid. If a quasiparticle is emitted, the new momentum will be $Q'+p$, where $Q'$ is the impurity momentum after collision, and the total energy becomes $E_f=Q'^2/(2m)+E_0+\epsilon(p)$. Imposing conservation of energy and momentum, the emission of excitation is allowed kinematically only when

\begin{equation}
    \frac{Q^2}{2m}\geq\frac{(Q-p)^2}{2m}+\epsilon(p)
\end{equation}
or equivalently,
\begin{equation}
    \epsilon(p)\leq\frac{p}{2m}(2Q-p).
    \label{eqn:parabola}
\end{equation}
The right hand side of Eq.~\eqref{eqn:parabola} describes a concave parabola that intersects the $p$-axis at $p\!=\!0$ and $p\!=\!2Q$, with an initial slope set by the impurity velocity $Q/m$. For an impurity mass equal to or larger than that of host particles, this parabola lies entirely below the lower edge of the excitation spectrum for $Q<k_\text{F}$ (Fig.~\ref{fig:multispectrum}), intersecting it only at $p=2\,k_\text{F}$ when $Q=k_\text{F}$. As a result, for impurity momenta below $k_\text{F}$, this kinematic argument predicts dissipationless motion.

\begin{figure}
    \centering
\includegraphics[trim={0cm 0cm 0cm 0cm},clip,width=\columnwidth]{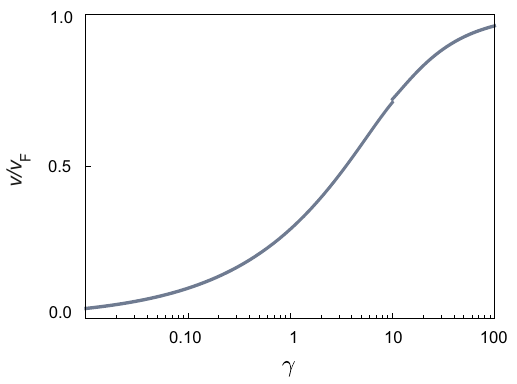}
\caption{ \textbf{Interaction dependence of the speed of sound.} Speed of sound $v$ (in units of the Fermi velocity $v_{F}$) in a 1D Bose gas as a function of the interaction parameter $\gamma$.}
	\label{fig:Sup_mat_sound}
\end{figure}

\begin{figure*}[t!]
    \centering
\includegraphics[trim={0cm 0cm 0cm 0cm},clip,width=\textwidth]{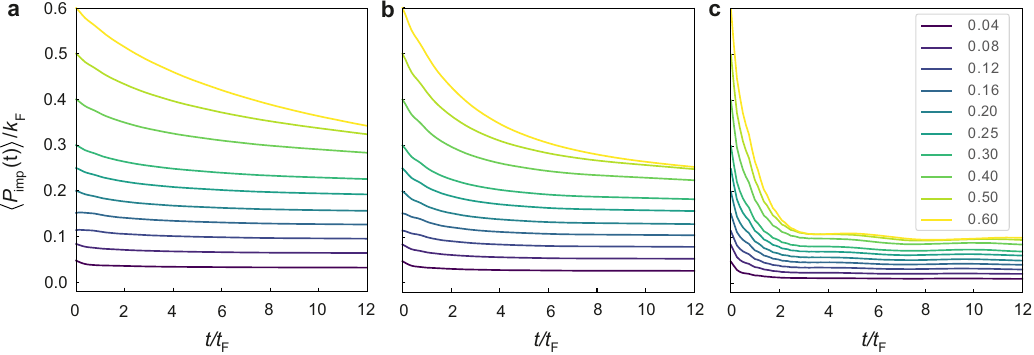}
\caption{ \textbf{Evolution of the mean impurity momentum.} \textbf{a}, \textbf{b}, and \textbf{c}, Numerical results for the mean impurity momentum as a function of evolution time $t/t_\text{F}$ for interaction parameters $(\gamma_i,\gamma) = (1.9,5.1), (3.5,7.4)$, and $(11.5,19.4)$, respectively. Initial impurity momenta $Q/k_\text{F}$ are indicated in the right panel.}
	\label{fig:Sup_mat_sim}
\end{figure*}

When considering polaron formation on top of this process, the impurity can form an entangled state with the quantum liquid. This is what we have confirmed experimentally in our work, and here we give some additional theoretical details. The energy of the free impurity of mass $m$ is related to its momentum $p$ b:
\begin{equation}
\varepsilon_\mathrm{imp} = \frac{p^2}{2m}. \label{freemagnondisp}
\end{equation}
This dispersion is shown on top of the energy spectrum of the 1D gas of particles of mass $m$ in Fig.~\ref{fig:disprsionmp}(a), as well as in Fig.~\ref{fig:Fig_1} of the main text. The impurity dispersion lies below the edge of the excitation spectrum of the 1D gas at a sufficiently small $k$, because the former is quadratic, Eq.~\eqref{freemagnondisp}, and the latter is linear around $k=0$, Eq.~\eqref{lindisp}. The excitation spectrum of a system containing an impurity that interacts with the 1D gas is shown in Fig.~\ref{fig:disprsionmp}(b). Notably, the edge of the spectrum has the same functional form at low energy and momentum as given by Eq.~\eqref{freemagnondisp} but with a renormalized mass:
\begin{equation}
\varepsilon_\text{pol} = \frac{p^2}{2m_\text{pol}} \qquad \text{for} \qquad p \ll \hbar k_\text{F}. \label{polarondisp}
\end{equation}
This shape of the edge represents a polaron, describing an impurity ``dressed'' by particle-hole excitations of the gas (the polaron mass $m_\text{pol}$ is typically greater than the bare impurity mass $m$ due to this dressing). The polaron state (which encompasses the moving impurity) has a lower energy than any phonon state (of gas particles without the impurity) at a small but non-zero momentum. Hence, a finite mass impurity can have a nonzero velocity in the steady state of a 1D interacting quantum system. We stress that this result is valid regardless of the details of the interaction potential in the host gas and between the host gas and the impurity particle.

\section{Theoretical model for the host gas and numerical simulations}

We model the host gas using the Lieb-Liniger Hamiltonian, which forms part of the full Hamiltonian~\eqref{eq:1} used in the main text. The Hamiltonian reads~\cite{LiebExact1963}

\begin{equation}
\hat{H}_\mathrm{LL}(g_\mathrm{1D},\{z_n\}) = \sum_{n=1}^N \frac{\hat{p}_n^2}{2m} + g_\mathrm{1D} \sum_{1\le n<n^\prime\le N} \delta(z_n-z_{n^\prime}), \label{eq:LLH}
\end{equation}
where particles of mass $m$ interact via a $\delta$-potential of strength $g_\mathrm{1D}$. We consider a homogeneous system with periodic boundary conditions on a ring of circumference $L$, and focus on the regime of large particle number $N$ at a fixed density $n_\text{1D}=N/L$. The interaction strength is characterized by the dimensionless parameter
\begin{equation}
\gamma=\frac{m g_\mathrm{1D}}{\hbar^2 n_\text{1D}}.
\label{gammadef}
\end{equation}
The excitation spectrum of the Lieb–Liniger gas is bounded from below by the minimal excitation energy $\varepsilon_\text{min}(k)$, as illustrated in Fig.~\ref{fig:multispectrum}. The function $\varepsilon_\text{min}(k)$ can be obtained by numerically solving the integral equations from Ref.~\cite{LiebExact1963}. The corresponding sound velocity, shown in Fig.~\ref{fig:Sup_mat_sound}, follows from Eq.~\eqref{vLLsound}.

\begin{figure*}[t!]
    \centering
\includegraphics[trim={0cm 0cm 0cm 0cm},clip,width=\textwidth]{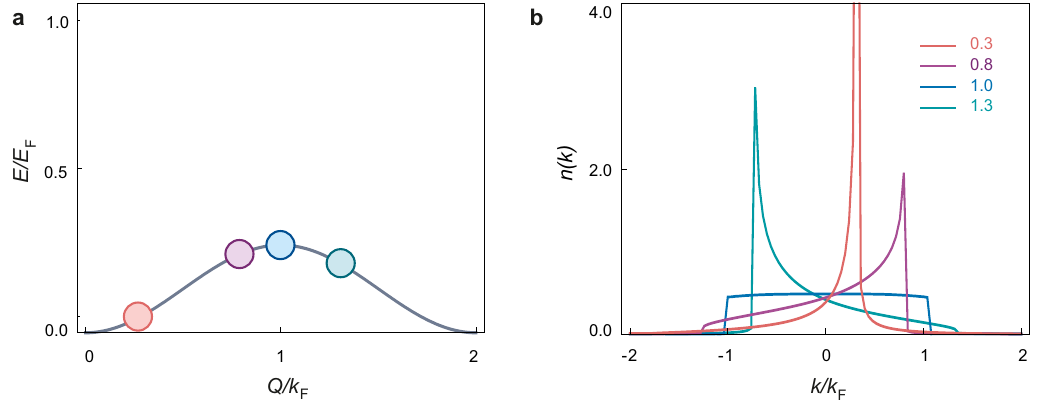}
\caption{ \textbf{Impurity momentum distribution in the polaron state.}  { \bf{a}}, Polaron spectrum for $\gamma_i=10$ assuming the host gas in Tonks-Girardeau limit. {\bf{b}}, Impurity momentum distribution $n(k)$ for total momentum $Q/k_\text{F} = 0.3,\, 0.8,\,1$ and $1.3$ along the excitation spectrum as indicated.}
	\label{fig:Sup_mat_Polaron}
\end{figure*}

In our numerical simulations, we discretize space and consider $N=40$ particles in a system of size $L=400\,d$, where $d$ is the lattice constant, and represent the wave function as a Matrix Product State (MPS)~\cite{Hauschild2018, Knap2014Flutter}. We (i) solve for the ground state in the absence of an impurity using the Density Matrix Renormalization Group (DMRG), (ii) insert an impurity with momentum $Q$, and (iii) performed time evolution with the Time Evolved Block Decimation (TEBD) to track the dynamics of the system~\cite{Hauschild2018}. Convergence of the results with bond dimension $\chi$ is checked. We typically use a bond dimension between 256 and 512. Additional examples for time traces of the impurity momentum are shown in Fig.~\ref{fig:Sup_mat_sim}. This data is used to determine the late time momentum $Q_f$ of the impurity shown in Fig. 5 of the main text.

\section{Impurity momentum distribution in the polaron state}
Figure~\ref{fig:Sup_mat_Polaron}\textbf{a} shows the calculated polaron spectrum of an impurity interacting with a Tonks-Girardeau host gas with impurity-host interaction strength $\gamma_i\!=\!10$. The impurity momentum distributions $n(k)$ in these polaron states are shown in Fig.~\ref{fig:Sup_mat_Polaron}\textbf{b} for different total momentum $Q/k_\text{F}$~\cite{Gamayun2020}. For both $Q<k_\text{F}$, the momentum distribution is asymmetric with a peak around $k\approx Q$. At $Q\!=\!k_\text{F}$, the distribution is the same as a system of free fermions. For $Q>k_{F}$, the distribution is again asymmetric, but now with a peak around $k\approx (2k_{F}-Q)$. In the limit of $\gamma_i \to \infty$, the impurity momentum distribution becomes identical to that of a system of 1D hardcore anyons~\cite{Gamayun2020,Santachiara2007,Dhar2025}.

\end{document}